\documentclass{aa}
\usepackage{graphicx}
\usepackage{txfonts}

\def\pmri{\object{1RXS~J001442.2+580201}}
\def\pmriii{\object{1RXS~J013106.4+612035}}
\def\pmriv{\object{1RXS~J042201.0+485610}}
\def\pmrvii{\object{1RXS~J062148.1+174736}}
\def\pmrix{\object{1RXS~J072259.5$-$073131}}
\def\pmrx{\object{1RXS~J072418.3$-$071508}}

\def\spmri{J001442.2+580201}
\def\spmriii{J013106.4+612035}
\def\spmriv{J042201.0+485610}
\def\spmrvii{J062148.1+174736}
\def\spmrix{J072259.5$-$073131}
\def\spmrx{J072418.3$-$071508}

\def\inv{$^{-1}$}

\newcommand\mconc[1]{\multicolumn{1}{c}{#1}}
\newcommand\mctwc[1]{\multicolumn{2}{c}{#1}}
\newcommand\mcthc[1]{\multicolumn{3}{c}{#1}}
\newcommand\mcfoc[1]{\multicolumn{4}{c}{#1}}

\newcommand\mcsic[1]{\multicolumn{6}{c}{#1}}

\newcommand\mcnic[1]{\multicolumn{9}{c}{#1}}

\newcommand\mconcl[1]{\multicolumn{1}{|l}{#1}}

\newcommand\mcnicl[1]{\multicolumn{9}{|c}{#1}}

\begin{document}

\title{EVN and MERLIN observations of microquasar candidates \\ at low galactic latitudes}
%\subtitle{II. EVN and MERLIN observations}

\author{M. Rib\'o\inst{1}
\and E. Ros\inst{2}
\and J.~M. Paredes\inst{1}
\and M. Massi\inst{2}
\and J. Mart\'{\i}\inst{3}
}

\offprints{M. Rib\'o, \\
\email{mribo@am.ub.es}}

\institute{Departament d'Astronomia i Meteorologia, Universitat de Barcelona, Av. Diagonal 647, 08028 Barcelona, Spain
\and Max-Planck-Institut f\"ur Radioastronomie, Auf dem H\"ugel 69, 53121 Bonn, Germany
\and Departamento de F\'{\i}sica, Escuela Polit\'ecnica Superior, Universidad de Ja\'en, Virgen de la Cabeza 2, 23071 Ja\'en, Spain
}

\date{Received 2 August 2002 / Accepted 27 August 2002}

\abstract{
In an attempt to increase the number of known microquasars, Paredes et~al.
(\cite{paredes02}) have presented a long-term project focused on the search
for new objects of this type. They performed a cross-identification between
X-ray and radio catalogs under very restrictive selection criteria for sources
with $|b|<5\degr$, and obtained a sample of 13 radio-emitting X-ray sources.
Follow-up observations of 6 of these sources with the VLA provided accurate
coordinates, which were used to discover optical counterparts for all of them.
We have observed these six sources with the EVN and MERLIN at 5~GHz. Five of
the six objects have been detected and imaged, presenting different
morphologies: one source has a two-sided jet, three sources have one-sided
jets, and one source is compact. With all the presently available information,
we conclude that two of the sources are promising microquasar candidates in
our Galaxy.
\keywords{
X-rays: binaries --
radio continuum: stars 
} 
}

\maketitle

\section{Introduction \label{sec:intro}}

Microquasars are stellar-mass black holes or neutron stars that mimic, on
smaller scales, many of the phenomena seen in AGN and quasars. Microquasars
have been found in X-ray binary systems, where a compact object accretes
matter from a companion star. Radio Emitting X-ray Binaries (REXBs) with
relativistic radio jets, like \object{SS~433}, \object{GRS~1915+105},
\object{GRO~J1655$-$40} or \object{Cyg~X-3}, are good examples of microquasars
(see Mirabel \& Rodr\'{\i}guez \cite{mirabel99} for a detailed review), while
other well known sources like \object{LS~I~+61~303} could turn out to be new
relativistic jet sources (Massi et~al. \cite{massi02}). With the recent
addition of \object{LS~5039} (Paredes et~al. \cite{paredes00}),
\object{Cyg~X-1} (Stirling et~al. \cite{stirling01}) and
\object{XTE~J1550$-$564} (Hannikainen et~al. \cite{hannikainen01}) to the
microquasar group, the current known number of this kind of sources is 14,
among $\sim50$ REXBs within the $\sim280$ known X-ray binaries (Liu et~al.
\cite{liu00}; Liu et~al. \cite{liu01}). Recent studies of microquasars can be
found in Castro-Tirado et~al. (\cite{castro01}). 

As pointed out in Sect.~1 of Paredes et~al. (\cite{paredes02}, hereafter
Paper~I), the number of known microquasars remains still small, especially
when trying to study them from a statistical point of view. An interesting
aspect of microquasars is their possibility of being related to unidentified
high-energy $\gamma$-ray sources, as suggested by Paredes et~al.
(\cite{paredes00}). Moreover, some parameters like the jet velocity, seem to
be related to the mass of the compact object (i.e., with its potential well).
Nevertheless, the lack of meaningful statistical studies because of the small
population of microquasars with known jet velocities, prevents any definitive
statements of this kind being made. Therefore, it is worth searching for new
microquasars in order to increase the known population.

\begin{table*}
\begin{center}
\caption[]{MERLIN positions for the five detected target sources, obtained via phase-referencing. The uncertainties quoted for the target sources were provided by the {\sc jmfit} task in {\sc aips} (measuring in the phase-referenced image) and do not include systematic errors. The positions of all phase-reference calibrators except \pmrx\ have an accuracy better than 1~mas.}
\label{table:merlinpos}
\begin{tabular}{@{}lr@{$^\mathrm{h}$}r@{$^\mathrm{m}$}r@{$\rlap{.}^\mathrm{s}$}l@{~}r@{\degr}r@{\arcmin}r@{$\rlap{.}\arcsec$}l|lr@{$^\mathrm{h}$}r@{$^\mathrm{m}$}r@{$\rlap{.}^\mathrm{s}$}l@{~}r@{\degr}r@{\arcmin}r@{$\rlap{.}\arcsec$}l@{}}
\hline \hline \noalign{\smallskip}
        \mcnic{Target sources}                  & \mcnicl{Phase-reference calibrators} \\
1RXS name & \mcfoc{$\alpha$ (J2000.0)} & \mcfoc{$\delta$ (J2000.0)}
          & \mconcl{Source name}             &  \mcfoc{$\alpha$ (J2000.0)} & \mcfoc{$\delta$ (J2000.0)} \\
\noalign{\smallskip} \hline \noalign{\smallskip}
\spmri\   &     00&14&42&12822 &     +58&02&01&2460 & J0007+5706$^\mathrm{a}$ & 00&07&48&47110   & +57&06&10&4540 \\
          &\mctwc{~} & $\pm$0&00013 & \mctwc{~} & $\pm$0&0022 &                       & \mcfoc{~}          & \mcfoc{~} \\
\spmriii\ &     01&31&07&23210 &     +61&20&33&3752 & J0147+5840$^\mathrm{a}$ & 01&47&46&54380   & +58&40&44&9750 \\
          &\mctwc{~} & $\pm$0&00014 & \mctwc{~} & $\pm$0&0016 &                       & \mcfoc{~}          & \mcfoc{~} \\
%\spmriv\ &     04&22&00&5244  &     +48&56&03&634  & NRAO~150$
\spmrvii\ &     06&21&47&75264 &     +17&47&35&0818 & J0630+1738$^\mathrm{b}$ & 06&30&07&25870   & +17&38&12&9300 \\
          &\mctwc{~} & $\pm$0&00006 & \mctwc{~} & $\pm$0&0017 &                       & \mcfoc{~}          & \mcfoc{~} \\
\spmrix\  &     07&22&59&68188 &   $-$07&31&34&8009 & \pmrx$^\mathrm{c}$    & 07&24&17&2912      & $-$07&15&20&339 \\
          &\mctwc{~} & $\pm$0&00011 & \mctwc{~} & $\pm$0&0022 &                       & \mcfoc{~}          & \mcfoc{~} \\
\spmrx$^\mathrm{d}$   
          &     07&24&17&2912  &   $-$07&15&20&339  & J0730$-$116           & 07&30&19&1125      & $-$11&41&12&601 \\
          &\mctwc{~} & $\pm$0&0007  & \mctwc{~} & $\pm$0&010  &                       & \mcfoc{~}          & \mcfoc{~} \\
\noalign{\smallskip} \hline
\end{tabular}
\begin{list}{}{
}
\item[$^{\rm a}$] Position from Patnaik et~al. (\cite{patnaik92}).
\item[$^{\rm b}$] Position from Browne et~al. (\cite{browne98}).
\item[$^{\rm c}$] This position was obtained from VLA observations (listed in the line below).
\item[$^{\rm d}$] The entries in this line correspond to VLA observations (see Paper~I).
\end{list}
\end{center}
\end{table*}
%------------------------------------------------------------------------------

In this context, the sources to search for are REXBs, which are in fact
microquasar candidates. To this end, a cross-identification between the X-ray
ROSAT all sky Bright Source Catalog (RBSC) (Voges et~al. \cite{voges99}) and
the NRAO VLA Sky Survey (NVSS) (Condon et~al. \cite{condon98}) was made for
sources with $|b|<5\degr$ under very restrictive selection criteria, and the
obtained results have been presented in Paper~I. A sample containing 13
sources was obtained, and 6 of them were observed at radio wavelengths in
order to obtain radio spectra and their variability, as well as accurate radio
positions, which allowed the authors to discover the corresponding optical
counterparts for all of them. At the end of this study, two of the sources,
namely \object{1RXS~J001442.2+580201} and \object{1RXS~J013106.4+612035} were
classified as promising microquasar candidates. The remaining four sources
were classified as weaker candidates for a number of reasons.
\object{1RXS~J042201.0+485610} shows a highly inverted spectrum at high radio
frequencies and an optical counterpart slightly extended.
\object{1RXS~J062148.1+174736} is an object extended in the optical.
\object{1RXS~J072259.5$-$073131} presents a one-sided radio jet at arcsecond
scales, supporting the possibility of being extragalactic. And finally,
\object{1RXS~J072418.3$-$071508} is a known quasar.

In this paper we present Very Long Baseline Interferometry (VLBI) observations
of these six sources, aimed at revealing possible jet-like features at
milliarcsecond scales. We describe the observations and the data reduction in
Sect.~\ref{sec:obs}, present the results and a discussion in
Sect.~\ref{sec:results}, and we summarize our findings in
Sect.~\ref{sec:summary}.

\section{Observations and data reduction \label{sec:obs}}

We observed the six sources studied in Paper~I simultaneously with the
Multi-Element Radio-Linked Interferometer Network (MERLIN) and the European
VLBI Network (EVN) on February 29th/March 1st 2000 (23:30--23:05~UT) at 5~GHz.
Since some of the target radio sources were faint, we scheduled the
observations introducing phase-reference calibrators with cycle times of
around 7~min (compatible with the expected coherence times). Apart from the
calibrators presented in Table~\ref{table:merlinpos}, \object{4C~61.02} was
used as calibrator for \pmriii, and \object{TXS~0422+496} and
\object{NRAO~150} for \pmriv. We also observed the fringe-finder
\object{DA~193} and the MERLIN flux density calibrator \object{3C~286}. Single
dish flux density measurements were carried out with the MPIfR 100~m antenna
in Effelsberg, Germany.

\subsection{MERLIN \label{subsec:obsmerlin}}

MERLIN is a connected radio interferometer across England, with baselines
reaching up to 217~km length. This array observed with 2-bit sampling at dual
polarisation with two blocks of 16 channels, each channel of 1~MHz bandwidth.
We analysed the left hand circular polarisation data excluding one channel at
both edges of the band, yielding a final bandwidth of 14~MHz. The correlator
integration time was of 4~s.
%Some antennas did not observe during short periods due to strong winds. 

The MERLIN data reduction was carried out at Jodrell Bank Observatory, using
standard procedures within the Astronomical Image Processing System ({\sc
aips}, developed and maintained by the US National Radio Astronomy
Observatory). We did not detect \pmriv. All other sources were detected, and
accurate positions for the target radio sources were obtained via
phase-referencing. These positions, presented in Table~\ref{table:merlinpos},
were used later as {\it a priori} information for the VLBI correlation. The
position given in Table~\ref{table:merlinpos} for the quasar \pmrx\ was
obtained from the VLA observations presented in Paper~I. The position of
\pmrix\ is deduced from the phase-reference offset relative to \pmrx\ provided
by MERLIN.

To image the sources, we averaged the data in frequency and exported them to
be processed into the difference mapping software {\sc difmap} (Shepherd
et~al. \cite{shepherd94}), where we time-averaged the data in 32~s bins after
careful editing.

\subsection{EVN \label{subsec:obsevn}}

The EVN observations were performed with the following
array (name, code, location, diameter): 
Effelsberg, EB, Germany, 100~m;
Jodrell Bank, JB, U.K., 25~m;
Cambridge, CM, U.K., 32~m;
Westerbork, WB, The Netherlands, 14$\times$25~m;
Medicina, MC, Italy, 32~m;
Noto, NT, Italy, 32~m;
Shanghai, SH, China, 25~m;
Toru\'n, TR, Poland, 32~m;
and
Onsala85, ON, Sweden, 25~m.
%,array described in Table~\ref{table:evnarray}, 
Data were recorded in MkIV mode with 2-bit sampling at 256~Mbps with left hand
circular polarization. A bandwidth of 64~MHz was used, divided into 8
intermediate frequency (IF) bands. 

The data were processed at the EVN MkIV correlator at the Joint Institute for
VLBI in Europe (JIVE), in Dwingeloo, The Netherlands. The correlator
integration time was of 4~s. A first post-processing analysis was also carried
out at JIVE. The data were processed using {\sc aips}. A first {\it a priori}
visibility amplitude calibration was performed using antenna gains and system
temperatures measured at each antenna. The fringe fitting ({\sc fring}) of the
residual delays and fringe rates was performed for all the radio sources. No
fringes were found for \pmriv\ and \object{4C~61.02}. Fringes for many
baselines were missing for \pmri, \pmrvii\ and \object{TXS~0422+496}.

To improve the fringe detection on all baselines for \pmri\ and \pmrvii\ we
used the delay, rate, and phase solutions from their corresponding
phase-reference calibrators (\object{J0007+5706}, 1\degr18\arcmin\ separation,
and \object{J0630+1738}, 1\degr59\arcmin\ separation) 
%, see Table~\ref{table:obsscheme}) 
and interpolated them to the target sources using the {\sc aips} task {\sc
clcal}. We fringe-fitted the target sources again using narrower search
windows and obtained solutions for all baselines. A similar attempt on \pmriv\
(with respect to \object{TXS~0422+486} and \object{NRAO~150}) was unfruitful.
Effelsberg was used as reference antenna throughout the {\sc aips} data
reduction process. 

We then averaged the data in frequency and exported them to be imaged and
self-calibrated in {\sc difmap}. The {\it a priori} visibility amplitude
calibration was not sufficient to reliably image the weakest radio sources. We
improved that by first imaging in {\sc difmap} the calibrator sources
\object{J0007+5706}, \object{J0147+5840}, and \object{J0630+1738}, with
appropriate amplitude self-calibration. We deduced correction factors for each
antenna, these being consistent for the three radio sources within 2\%.
%The factors were: EB:0.98, JB:1.10, CM:0.78, WB:1.18, MC:0.87, NT:1.12,
%SH:1.03, TR:0.98 and ON:0.97. 
We corrected the amplitude calibration back in {\sc aips} 
%using the task {\sc sncor} ({\sc optype 'mula'}) for all the radio sources 
and exported the data again into {\sc difmap}, where the final imaging was
performed after editing and averaging of the visibilities in 32~s blocks.

\subsection{Combining EVN and MERLIN \label{subsec:evn+merlin}}

The EVN and MERLIN arrays have one common baseline, between JB and CM, which allows to combine both data sets and map them together.
%Combining data sets from both arrays allows us to reach $(u,v)$ resolution
%ranges from 0.04~M$\lambda$ (MK2-Tabley) to 140~M$\lambda$ (NT-SH) at 5~GHz. 
We processed the data within {\sc aips} in order to combine both arrays. The
B1950.0 $(u,v)$ coordinates of the MERLIN data had to be corrected to the ones
of the EVN for the same reference system (J2000.0) with {\sc uvfix}. Then, the
MERLIN data were self-calibrated with the EVN images (see below), and the
phase solutions were limited to the longest MERLIN baselines. The MERLIN data
were imaged and the peak-of-brightness of both data sets were checked to be
similar. As a next step, the EVN data were averaged in frequency to correspond
to the MERLIN data. The {\sc aips} headers of both data sets were modified
conveniently to match together, and the weighting of both data sets was also
modified to be equal with {\sc wtmod}. Finally, both data sets were
concatenated (using {\sc dbcon}) and exported to {\sc difmap} to be imaged
with different data weighting in $(u,v)$ distance (tapering) after time
averaging in 32~s bins.

\subsection{Flux density measurements at the 100~m antenna in Effelsberg
\label{subsec:effelsberg}}

%To complement the amplitude calibration and obtain additional information on
%the radio sources, 
We interleaved cross-scans (in azimuth and elevation) with the 100~m
Effelsberg antenna to measure the radio source flux densities (A.~Kraus,
private communication). We fitted a Gaussian function to the flux-density
response for every cross-scan, and we averaged the different Gaussians. We
linked the flux density scale by observing primary calibrators such as
\object{3C~286}, \object{3C~48}, or \object{NGC~7027} (see e.g. Kraus
\cite{kraus97}; Peng et~al. \cite{peng00}). We list the single dish flux
density measurements in the second column of Table~\ref{table:param}. The flux
density values for the main VLBI calibrators were of
%0.177$\pm$0.004~Jy for 0803+04,
%0.538$\pm$0.011~Jy for 2327+09,
%16.1$\pm$0.2~Jy for 3C~123,
%3.72$\pm$0.07~Jy for 3C~138,
%7.89$\pm$0.08~Jy for 3C~147,
%4.31$\pm$0.06~Jy for 3C~196,
%40.4$\pm$0.6~Jy for 3C~273,
%20.4$\pm$0.4~Jy for 3C~279,
$7.5\pm0.1$~Jy for 3C~286,
%6.57$\pm$0.08~Jy for 3C~295,
%5.5$\pm$0.1~Jy for 3C~48,
%20.6$\pm$0.3~Jy for 3C~84,
and
$5.9\pm0.2$~Jy for DA~193.
%0.72$\pm$0.06 for J1823-12,
%5.52$\pm$0.08 for NGC~7027.
%Those values helped us to calibrate both 
%the MERLIN and the EVN observations.

\section{Results and discussion \label{sec:results}}

We present all the imaging results in Figs.~\ref{fig:pmr1}--\ref{fig:pmr10},
and the image parameters in Table~\ref{table:param}. The total flux density
values for the different images diverge from each other and from the single
dish measurements, due to the amplitude self-calibration process in all cases.
Therefore, those values should be considered with care. The minimum contours
in the images are those listed as $S_\mathrm{min}$ in Table~\ref{table:param},
while consecutive higher contours scale with $3^{1/2}$. Here follows a
detailed discussion on each source.

%------------------------------------------------------------------------------
\begin{table*}
\begin{center}
\caption[]{Flux densities and parameters used to produce the images in Figs.~\ref{fig:pmr1}--\ref{fig:pmr10}.}
\label{table:param}
\begin{footnotesize}
\begin{tabular}{@{}lr@{$\,\pm\,$}lcr@{$\,\times\,$}lcccc@{}}
\hline \hline \noalign{\smallskip}
          & \mctwc{Single dish}    & \mconc{Array$^\mathrm{a}$}     
                                              &\mcsic{~} \\
1RXS name & \mctwc{$S_\mathrm{EB}$}& \mconc{(taper FWHM)}   
                                              & \mctwc{beam~size} & P.A. & $S_\mathrm{tot}$ & $S_\mathrm{peak}$ & $S_\mathrm{min}$ \\
          &\mctwc{[mJy]}
                                   & \mconc{[M$\lambda$]}
                                              & [mas] & [mas] 
                                                              & \mconc{[$\degr$]} 
                                                                         & [mJy]
                                                                                          & [mJy~beam\inv] 
                                                                                                        & [mJy~beam\inv] \\
\noalign{\smallskip} \hline \noalign{\smallskip}
\spmri\   &   6.5&0.5$^\mathrm{b}$ & M        & 57   & 51     &      ~~\,37 &      ~~~6.2 &      ~~~5.8 & 0.1  \\
          & \mctwc{~}              & E+M (10) & 7.7  & 7.2    &     ~\,$-$4 &     ~\,10.1 &      ~~~9.9 & ~\,0.18 \\
          & \mctwc{~}              & E        & 1.77 & 0.86   &       $-$20 &     ~\,11.5 &      ~~~7.0 & ~\,0.15 \\
\noalign{\smallskip}
\spmriii\ &  20.1&0.6              & M        & 71   & 39     &       $-$63 &     ~\,17.5 &     ~\,17.9 & 0.5  \\
          & \mctwc{~}              & E+M (15) & 7.8  & 6.2    &       $-$75 &     ~\,19.2 &     ~\,18.7 & 0.8  \\
          & \mctwc{~}              & E        & 1.04 & 0.99   &     ~\,$-$1 &     ~\,17.6 &     ~\,11.6 & 0.4  \\
\noalign{\smallskip}
\spmriv\  & \mctwc{$<5$}           & ---      & \mctwc{---}   & \mconc{~\,---} 
                                                                            & \mconc{~~\,---} 
                                                                                          & \mconc{~~\,---} 
                                                                                                        & \mconc{~~\,---} \\
\noalign{\smallskip}
\spmrvii\ &   7.1&0.7$^\mathrm{b}$ & M        & 88   & 39     &      ~~\,24 &      ~~~5.8 &      ~~~5.6 & 0.2  \\
          & \mctwc{~}              & E+M (8)  & 13.1 & 11.1   &       $-$24 &      ~~~6.0 &      ~~~6.8 & 0.3  \\
          & \mctwc{~}              & E        & 8.8  & 4.1    &      ~~\,47 &      ~~~7.0 &      ~~~6.4 & 0.3  \\
\noalign{\smallskip}
\spmrix\  &  67.9&1.1              & M        & 115  & 66     &      ~~~\,4 &     ~\,66.0 &     ~\,62.9 & 0.7  \\
          & \mctwc{~}              & E+M (15) & 9.5  & 7.3    &       $-$61 &     ~\,52.1 &     ~\,41.9 & 0.9  \\
          & \mctwc{~}              & E        & 5.74 & 1.12   &      ~~\,11 &     ~\,46.9 &     ~\,36.2 & 0.9  \\
\noalign{\smallskip}
\spmrx\   & 282.2&4.1              & M        & 120  & 64    &      ~~~\,8 &       301.0 &       285.7 & 0.9  \\
          & \mctwc{~}              & E+M (10) & 11.7 & 9.8    &       $-$61 &       287.0 &       263.4 & 2.0  \\
          & \mctwc{~}              & E        & 5.73 & 1.02   &      ~~\,11 &       248.0 &       184.8 & 0.7  \\
\noalign{\smallskip} \hline
\end{tabular}
\end{footnotesize}
\begin{list}{}{
}
\begin{footnotesize}
\item[$^{\rm a}$] M: MERLIN. E+M: EVN+MERLIN, FWHM of the tapering function (weighting of visibilities) in parenthesis. E: EVN.
\item[$^{\rm b}$] Values with low SNR in the Gaussian fits.
\end{footnotesize}
\end{list}
\end{center}
\end{table*}
%------------------------------------------------------------------------------

%\subsection{Discussion on individual sources \label{subsec:discindiv}}

\subsection{\pmri\ and its two-sided jet \label{subsec:pmr1}}

\begin{figure*}[htpb]
\resizebox{\hsize}{!}{\includegraphics{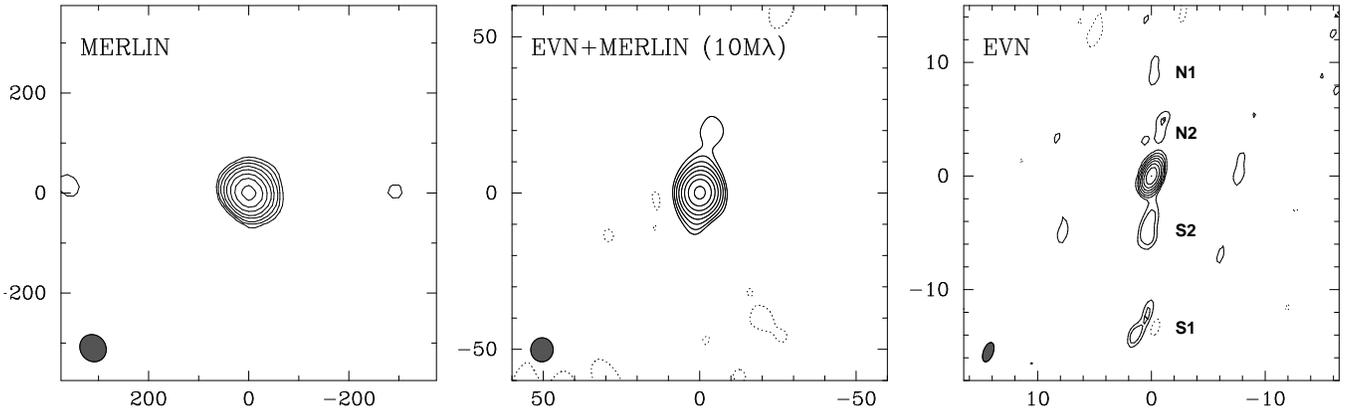}}
\caption{Images of \pmri\ using the different arrays and with the parameters given in Table~\ref{table:param}. The axes units are in mas. A $(u,v)$-tapering with a FWHM at 10~M$\lambda$ has been used to perform the combined EVN+MERLIN image. In the EVN image, N1, N2, S2 and S1 indicate the components discussed in the text.}
\label{fig:pmr1}
\end{figure*}

As can be seen in our images, shown in Fig.~\ref{fig:pmr1}, this source
appears point-like at MERLIN resolution, partially resolved in the tapered
EVN+MERLIN image and clearly resolved at EVN scales. In this last case, it
shows a two-sided jet-like structure roughly in the north-south direction,
with brighter components towards the south. The trends are visible in the
closure phases, giving us confidence that the structure observed is not a
consequence of sidelobes or imaging artifacts. In the tapered EVN+MERLIN
images, the structure extends up to 20--30~mas outside of the core, and more
clearly towards the north. This discrepancy could be due to calibration
problems.

%------------------------------------------------------------------------------
%------------------------------------------------------------------------------

Model fitting of the EVN visibilities with circular Gaussians provides a
parametrization of the inner structure. Five components reproduce the
visibilities. The central one has 7.8~mJy, with a FWHM of 0.4~mas. Towards the
north, one component (N2) of 0.6~mJy at 3.6~mas (P.A.~$-10\degr$, FWHM
0.7~mas) and another one (N1) of 0.5~mJy at 8.6~mas (P.A.~$2\degr$, extended
over 3~mas) are needed. Brighter components are present southwards, one (S2)
of 1.1~mJy at 5.0~mas (P.A.~$177\degr$, FWHM of 0.8~mas) and the other one
(S1) of 0.4~mJy at 13.7~mas (P.A.~$177\degr$, FWHM below 0.3~mas).

If we assume that components S1 and N1 correspond to a pair of plasma clouds
ejected at the same epoch near the compact object and perpendicularly to the
accretion disk, we can estimate some parameters of the jets by using the
following equation:
\begin{equation}
\beta\cos\theta={{\mu_{\rm a}-\mu_{\rm r}}\over{\mu_{\rm a}+\mu_{\rm r}}}={{d_{\rm a}-d_{\rm r}}\over{d_{\rm a}+d_{\rm r}}}~~,
\label{eqdist}
\end{equation}
$\beta$ being the velocity of the clouds in units of the speed of light,
$\theta$ the angle between the direction of motion of the ejecta and the line
of sight and $\mu_{\rm a}$ and $\mu_{\rm r}$ the proper motions of the
approaching and receding components, respectively (Mirabel \& Rodr\'{\i}guez
\cite{mirabel99}). Although we do not know the epoch of ejection of the
clouds, we can cancel the time variable by using the relative distances to the
core $d_{\rm a}$ and $d_{\rm r}$, as expressed in Eq.~\ref{eqdist}. Since both
variables, $\beta$ and $\cos\theta$, take values between 0 and 1, it is clear
that knowing $\beta\cos\theta$ allows us to compute a lower limit for the
velocity ($\beta_{\rm min}$) and an upper limit for the angle ($\theta_{\rm
max}$). The same applies for the S2 and N2 components. In
Table~\ref{table:components} we list the positions of the components obtained
from model fitting, together with the derived values from $\beta_{\rm min}$
and $\theta_{\rm max}$ for each one of the pairs. The slightly different
results obtained using pair 1 or 2, could be due to the fact that the position
for the S1 component obtained with model fitting happens to be at the lower
part of this elongated component, hence increasing $\beta\cos\theta$, or to
intrinsic different velocities for each one of the pairs. Hereafter we will
use $\beta>0.20\pm0.02$ and $\theta<78\pm1\degr$. 

A similar approach to obtain the jet parameters of the source can be performed
thanks to the brightness asymmetry of the components using the following
equation (Mirabel \& Rodr\'{\i}guez \cite{mirabel99}):
\begin{equation}
\beta\cos\theta={\big({S_{\rm a}/{S_{\rm r}}}\big)^{1/(k-\alpha)}-1 \over \big({S_{\rm a}/{S_{\rm r}}}\big)^{1/(k-\alpha)}+1}~~,
\label{eqflux}
\end{equation}
where $S_{\rm a}$ and $S_{\rm r}$ are the flux densities of the approaching
and receding components, respectively, $k$ equals 2 for a continuous jet and 3
for discrete condensations, and $\alpha$ is the spectral index of the emission
($S_{\nu}\propto \nu^{+\alpha}$). However, the equation above is only valid
when the components are at the same distance from the core. If this is not the
case (i.e., $d_{\rm r}<d_{\rm a}$), the ratio $S_{\rm a}/S_{\rm r}$ will be
lower than the one that should be used in Eq.~\ref{eqflux} (because the flux
density decreases with increasing distance from the core). In consequence,
Eq.~\ref{eqflux} only allows us to obtain a lower limit for $\beta\cos\theta$.

%------------------------------------------------------------------------------
\begin{table}
\begin{center}
\caption[]{Model fitted positions of the components in the \pmri\ radio jets and obtained jet parameters.}
\label{table:components}
\begin{footnotesize}
\begin{tabular}{@{}ccr@{~~~~}|@{~~~~}cc@{}}
\hline \hline \noalign{\smallskip}
Comp. & Distance       & P.A.    & $\beta_{\rm min}$ & $\theta_{\rm max}$\\
      & [mas]          & [\degr] &                   & [\degr]\\
\noalign{\smallskip} \hline \noalign{\smallskip} 
N1    & ~\,$8.6\pm0.3$ &     2   & $0.23\pm0.02$     & $77\pm1$\\
N2    & ~\,$3.6\pm0.2$ & $-$10   & $0.16\pm0.02$     & $81\pm1$\\
S2    & ~\,$5.0\pm0.1$ &   177   & $0.16\pm0.02$     & $81\pm1$\\
S1    &   $13.7\pm0.3$ &   177   & $0.23\pm0.02$     & $77\pm1$\\
\noalign{\smallskip} \hline
\end{tabular}
\end{footnotesize}
\end{center}
\end{table}
%------------------------------------------------------------------------------

In order to use this approach we will consider $k=3$, because the components
seem to be discrete condensations, and $\alpha=-0.20\pm0.05$, according to the
overall spectral index reported in Paper~I, since we do not have spectral
index information of the components. The use of the flux densities obtained
after model fitting gives $\beta\cos\theta>0.09\pm0.04$ for the S2--N2 pair
and $\beta\cos\theta>-0.07\pm0.08$ for the S1--N1 pair. This last value is
certainly surprising, although it can be explained by the fact that the S1
flux density obtained after model fitting does not account for the total flux
of this plasma cloud. In fact, better estimates of the flux densities can be
obtained summing together the flux densities of the {\sc clean} components
obtained within each one of the four plasma clouds. This yields to
$\beta\cos\theta>0.13\pm0.05$ for the S2--N2 pair and
$\beta\cos\theta>0.17\pm0.02$ for the S1--N1 pair, in good agreement with the
values computed using the distances from the components to the core, shown in
Table~\ref{table:components}.

If we compare the VLA position reported in Paper~I with the MERLIN position in
Table~\ref{table:merlinpos} we can see that they are different. In fact,
taking into account the errors, the MERLIN$-$VLA position offsets can be
expressed as: $\Delta\alpha\cos\delta=16\pm10$~mas and
$\Delta\delta=27\pm10$~mas. Assuming that the difference in position is due to
intrinsic proper motions and considering the time span between both
observations, 224 days, we obtain:
$\mu_{\alpha\cos\delta}=26\pm16$~mas~yr$^{-1}$ and
$\mu_{\delta}=44\pm16$~mas~yr$^{-1}$. Although a proper motion of the source
would clearly indicate its microquasar nature, because no proper motion would
be detected in an extragalactic source, we must be cautious with this result.
First of all, the phase-reference sources where different in the VLA and
MERLIN observations, as well as the observing frequencies from which positions
were estimated. On the other hand, none of these results exceeds the $3\sigma$
value, preventing to state that a proper motion has been detected.

Summarizing, our results indicate that this source exhibits relativistic radio
jets with $\beta>0.20$ and, therefore, together with the results reported in
Paper~I, we consider \pmri\ as a very promising microquasar candidate.

\subsection{\pmriii\ and its one-sided jet \label{subsec:pmr3}}

\begin{figure*}[htpb]
\resizebox{\hsize}{!}{\includegraphics{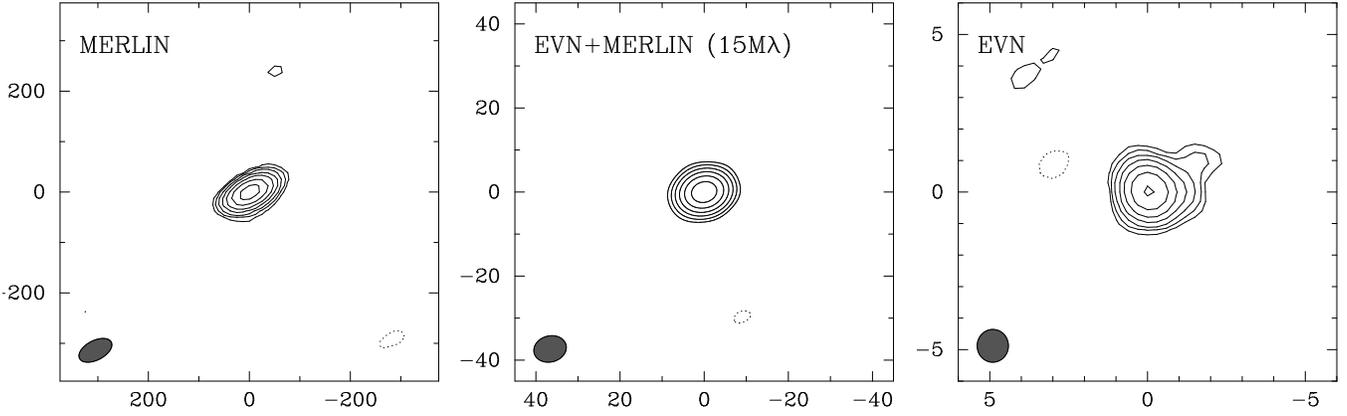}}
\caption{Images of \pmriii\ using the different arrays and with the parameters given in Table~\ref{table:param}. The axes units are in mas. A $(u,v)$-tapering with a FWHM at 15~M$\lambda$ has been used to perform the combined EVN+MERLIN image.}
\label{fig:pmr3}
\end{figure*}

We show in Fig.~\ref{fig:pmr3} the images obtained after our observations.
Although the source appears compact at MERLIN and EVN+MERLIN scales, there is
a weak one-sided radio jet towards the northwest in the EVN image. Model
fitting with circular Gaussian components can reproduce the observed
visibilities as follows: a central 15.4~mJy component with 0.64~mas FWHM, and
a northwest 2.1~mJy component with a FWHM of 0.74~mas, located at 1.8~mas in
P.A. $-73\degr$. As can be seen, this last component is located at a distance
of $\sim2$ times the beam size from the core.

%------------------------------------------------------------------------------
%------------------------------------------------------------------------------

Using the fact that we do not detect a counter-jet, we can use
Eq.~\ref{eqflux} replacing $S_{\rm r}$ with the $3\sigma$ level value. This,
of course, will only provide a lower limit to $\beta\cos\theta$, expressed as
follows:
\begin{equation}
\beta\cos\theta> {\big({S_{\rm a}/3\sigma}\big)^{1/(k-\alpha)}-1 \over \big({S_{\rm a}/3\sigma}\big)^{1/(k-\alpha)}+1}~~.
\label{eqsigma}
\end{equation}
Using $S_{\rm a}=2.1$~mJy, $3\sigma=0.30$~mJy (the $1\sigma$ value has been
taken as the root mean square noise in the image), $\alpha=-0.05\pm0.05$ (see
Paper~I), and $k=3$ to be consistent with the lowest limit, we obtain
$\beta\cos\theta>0.31\pm0.05$ ($\beta>0.31\pm0.05$ and $\theta<72\pm3$).
Hence, a lower limit of $\beta\geq0.3$ is obtained, pointing towards
relativistic radio jets as the origin of the elongated radio emission present
in the EVN image. Although the one-sided jet morphology at mas scales is found
mostly in extragalactic sources, it is also present in some galactic REXBs,
like \object{Cyg~X-3} (Mioduszewski et~al. \cite{mioduszewski01}) or
\object{LS~I~+61~303} (Massi et~al. \cite{massi01}). Hence, we cannot rule out
a possible galactic nature on the basis of the detected morphology.

%------------------------------------------------------------------------------
\begin{figure*}[htpb]
\resizebox{\hsize}{!}{\includegraphics{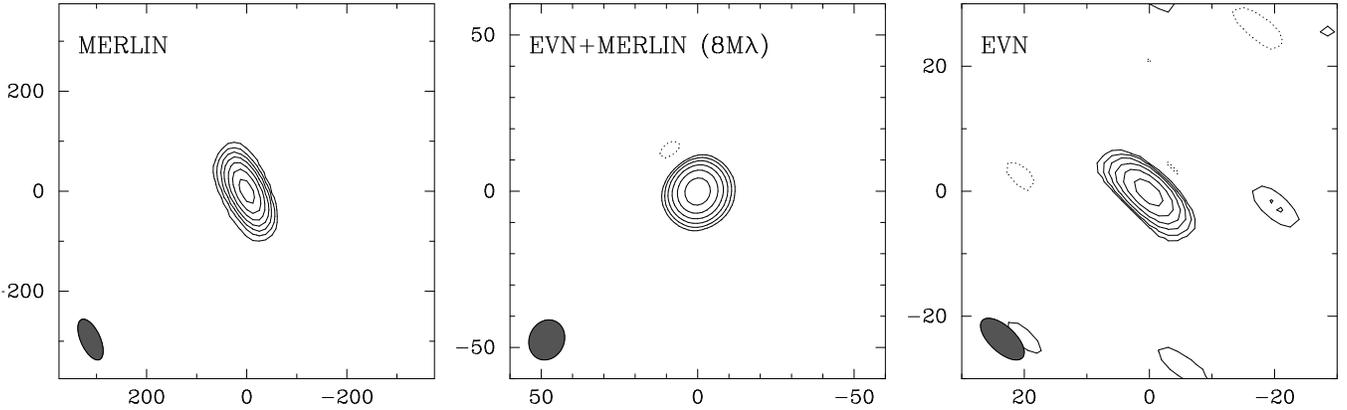}}
\caption{Images of \pmrvii\ using the different arrays and with the parameters given in Table~\ref{table:param}. The axes units are in mas. A $(u,v)$-tapering with a FWHM at 8~M$\lambda$ has been used to perform the combined EVN+MERLIN image.}
\label{fig:pmr7}
\end{figure*}
%------------------------------------------------------------------------------

As done for the previous source, we can compare the VLA position with the
MERLIN one, and find that they differ in $\Delta\alpha\cos\delta=39\pm10$~mas
and $\Delta\delta=-0.8\pm10$~mas. If the offsets are real, this would imply
$\mu_{\alpha\cos\delta}=64\pm16$~mas~yr$^{-1}$ and
$\mu_{\delta}=-1\pm16$~mas~yr$^{-1}$. Hence, it seems possible that we have
detected a proper motion in right ascension at a $4\sigma$ level. However, we
must be cautious since, as in the previous case, the phase-reference sources
and observing frequencies were different in each observation.

Overall, these results are indicative of relativistic radio jets, and hence,
together with the results reported in Paper~I, allow us to classify this
source as a promising microquasar candidate.

\subsection{\pmriv, a non-detected source \label{subsec:pmr4}}

This radio source was marginally seen in the cross-scans carried out in
Effelsberg. In fact, this is compatible with the low flux density of
$2.3\pm0.4$~mJy at 1.4~GHz listed in the NVSS. The source was not detected
with MERLIN or with the EVN, mainly due to problems with the phase-reference
sources, as pointed out in Sect.~\ref{subsec:obsevn}. In fact, as discussed in
Paper~I, VLA~A configuration observations showed a flux density of 0.4~mJy at
5~GHz and an inverted spectrum with a spectral index up to $\alpha=+1.6$,
suggesting thermal radio-emission resolved at higher angular resolutions.
However, we cannot exclude the possibility of having a highly variable source.

%------------------------------------------------------------------------------
\begin{figure*}[htpb]
\resizebox{\hsize}{!}{\includegraphics{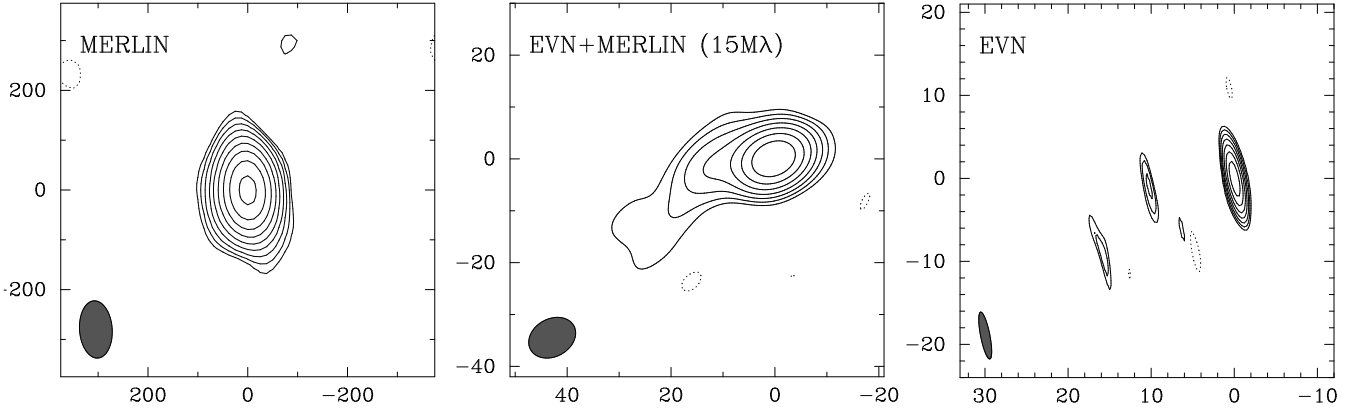}}
\caption{Images of \pmrix\ using the different arrays and with the parameters given in Table~\ref{table:param}. The axes units are in mas. A $(u,v)$-tapering with a FWHM at 15~M$\lambda$ has been used to perform the combined EVN+MERLIN image.}
\label{fig:pmr9}
\end{figure*}
%------------------------------------------------------------------------------

\subsection{\pmrvii, a compact source \label{subsec:pmr7}}

As can be seen in our images, shown in Fig.~\ref{fig:pmr7}, the radio source
is compact on all scales. In fact, model fitting of the EVN visibilities
converges to a point-like radio source (the FWHM of a circular Gaussian tends
to zero). We must note that when this source was observed it was below the
horizon in SH. Therefore, the obtained beam size for the EVN image is larger
than the ones obtained for the other sources, as can be seen in
Table~\ref{table:param}, hence providing lower angular resolution than in the
other cases. A comparison between the VLA and MERLIN positions reveals that
they are perfectly compatible within the errors, suggesting an extragalactic
nature or a small proper motion if it turns out to be a galactic microquasar.
Although the compactness of the source is not indicative of a galactic or an
extragalactic origin, the extended optical counterpart, reported in Paper~I,
suggests an extragalactic origin for this source. In fact, the detected radio
variability from 8 to 11~mJy within a week reported in Paper~I, is compatible
with the compactness in an extragalactic object (see IDV phenomenon, Wagner \&
Witzel \cite{wagner95}).

\subsection{\pmrix\ and its bent one-sided jet \label{subsec:pmr9}}

The images obtained at different resolutions are plotted in
Fig.~\ref{fig:pmr9}. The MERLIN image presents a compact structure with some
elongation eastwards, while the EVN and combined EVN+MERLIN images show a
clear one-sided jet towards the east, with a slight bent towards the south at
larger core separations. The closure phases clearly show that the source
departs from symmetry, with preferred emission to the east, both in the MERLIN
and the EVN data sets. Two distinct components are present in the EVN image,
at 9 and 17~mas from the compact core (P.A. of 89 and 113$\degr$,
respectively). Those components can be model fitted with elliptical Gaussians,
yielding flux densities of 3.7 and 2.8~mJy, respectively, for a core of
40.3~mJy.

%------------------------------------------------------------------------------
\begin{figure*}[htpb]
\resizebox{\hsize}{!}{\includegraphics{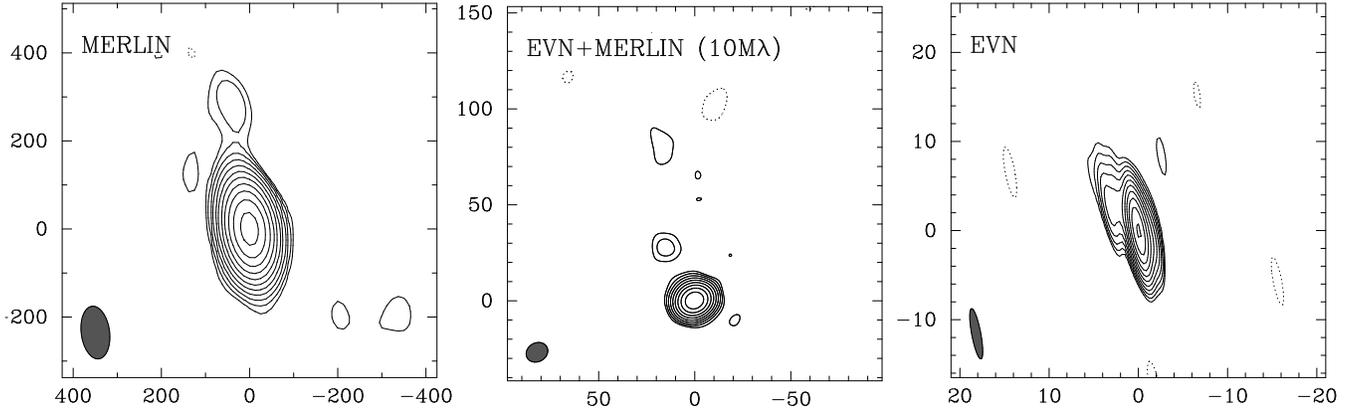}}
\caption{Images of the already identified quasar \pmrx\ using the different arrays and with the parameters given in Table~\ref{table:param}. The axes units are in mas. A $(u,v)$-tapering with a FWHM at 10~M$\lambda$ has been used to perform the combined EVN+MERLIN image.}
\label{fig:pmr10}
\end{figure*}
%------------------------------------------------------------------------------

Using again the fact that we do not detect a counter-jet, we can use
Eq.~\ref{eqsigma} with $S_{\rm a}=3.7$~mJy (the closest component to the core
in the EVN image), $3\sigma=0.54$~mJy (as previously done, the $1\sigma$ value
has been taken as the root mean square noise in the image),
$\alpha=-0.25\pm0.2$ (see Paper~I), and $k=3$ to be consistent with the
discrete nature of the components, to obtain $\beta\cos\theta>0.29\pm0.05$,
and hence $\beta>0.29\pm0.05$ and $\theta<73\pm3\degr$. Therefore, these
results point towards relativistic radio jets as the origin of the elongated
radio emission present in the images.

As pointed out for \pmriii, the one-sided jet morphology does not rule out a
microquasar nature for this object. However, the bending of the jet at such
small angular scales resembles the ones seen in blazars. In fact, as reported
in Paper~I, a one-sided arcsecond scale jet is also present in VLA~A
configuration observations at 1.4~GHz, an unusual feature in the already known
microquasars.

A comparison between the VLA and MERLIN positions reveals that they agree
within the errors, suggesting an extragalactic nature or a small proper motion
if it turns out to be a galactic microquasar.

Overall, these results are indicative of relativistic radio jets, although
this source shows characteristics more similar to blazars than to
microquasars.

\subsection{\pmrx, a quasar with a bent one-sided jet \label{subsec:pmr10}}

This radio source has recently been (March 2002) classified as a quasar in the
SIMBAD database, and it is not any more a microquasar candidate. It is
catalogued in the Parkes-MIT-NRAO (Griffith et~al. \cite{griffith94}) and the
Texas Survey (Douglas et~al. \cite{douglas96}). It is listed as
\object{PMN~J0724$-$0715} in the NED database, and is the source
\object{WGA~J0724.3$-$0715} in Perlman et~al. (\cite{perlman98}), who reported
a faint and quite broad H$\alpha$ emission line (rest-frame
$W_{\lambda}=30.3$~\AA, FWHM=4000~km~s$^{-1}$), and classified it as a Flat
Spectrum Radio Quasar (FSRQ) with $z=0.270$. Nevertheless, we have reported
here our observational results for this source, since it was a candidate when
we performed the observations. It presents (Fig.~\ref{fig:pmr10}) a one-sided
pc-scale jet oriented towards the northeast, changing from a P.A. of
$\sim50\degr$ at 5~mas from the core (EVN image) to $20\degr$ up to 200~mas,
at MERLIN scales.

Model fitting of the EVN visibilities with circular Gaussians reveals a
compact core (0.7~mas FWHM) with 245.4~mJy, and two distinct components, one
with 11.9~mJy at 6.2~mas (P.A. $46\degr$, 1.9~mas FWHM, present in the right
panel image of Fig.~\ref{fig:pmr10}) and another one with the size of the
beam, a flux density of 1.7~mJy at a distance of 29.6~mas in P.A. $27\degr$
(visible in the middle panel image of Fig.~\ref{fig:pmr10}).

%------------------------------------------------------------------------------
\begin{table*}
\begin{center}
\caption[]{Summary of the obtained results after the VLA and optical observations (Paper~I), and the EVN and MERLIN observations reported here. An asterisk indicates a non-expected behaviour for microquasars.}
\label{table:summary}
\begin{footnotesize}
\begin{tabular}{@{}ll@{\,}cl@{~\,}cl@{~\,}c@{~~}cl@{}}
\hline \hline \noalign{\smallskip}
1RXS name & \mctwc{VLA obs.}                & \mctwc{Optical obs.}    & \mcthc{EVN+MERLIN obs.}                              & Notes\\
          & Structure         & $\alpha$    & Structure    & $I$ mag. & Structure              & $\beta$   & $\theta[\degr]$ & \\
\noalign{\smallskip} \hline \noalign{\smallskip}
\spmri\   & compact           & $-0.2$      & point-like   & 19.9     & two-sided jet          & $>0.20$   & $<78$           & Promising candidate\\
\spmriii\ & compact           & $-0.1$      & point-like   & 17.9     & one-sided jet          & $>0.31$   & $<72$           & Promising candidate\\
\spmriv\  & compact           & ~\,$+1.6^*$ & extended$^*$ & 17.5     & not detected$^*$       & ---       & ---             & Thermal source ?\\
\spmrvii\ & compact           & $+0.1$      & extended$^*$ & 17.6     & compact                & ---       & ---             & Galaxy ?\\
\spmrix\  & one-sided jet$^*$ & $-0.2$      & point-like   & 16.8     & bent one-sided jet$^*$ & $>0.29$   & $<73$           & Blazar ?\\
\spmrx\   & compact           & $+0.1$      & point-like   & 17.2     & bent one-sided jet$^*$ & $>0.51$   & $<60$           & Quasar (FSRQ)\\
\noalign{\smallskip} \hline
\end{tabular}
\end{footnotesize}
\end{center}
\end{table*}
%------------------------------------------------------------------------------

Using again the fact that we do not detect a counter-jet, we can use
Eq.~\ref{eqsigma} with $S_{\rm a}=11.9$~mJy (the closest component to the
core), $3\sigma=0.45$~mJy, $\alpha=0.07\pm0.02$ (see Paper~I), and $k=3$ to be
consistent with the discrete nature of the components, to obtain
$\beta\cos\theta>0.51\pm0.04$. Hence, an upper limit of $\theta<60\pm3\degr$
and a lower limit of $\beta>0.51\pm0.04$ is obtained, pointing towards
relativistic radio jets as the origin of the elongated radio emission present
in the images of this already identified quasar.

\section{Summary \label{sec:summary}}

We have presented EVN and MERLIN observations of the six sources studied by
Paredes et~al. (\cite{paredes02}) in their search for microquasar candidates
at low galactic latitudes. The first one, namely \pmri, displays a two-sided
radio jet, which after analysis implies $\beta>0.20\pm0.02$ and
$\theta<78\pm1\degr$. \pmriii, displays a one-sided radio jet, requiring
$\beta>0.31\pm0.05$ and $\theta<72\pm3\degr$. The third one, namely \pmriv,
was not detected due to its low flux density and/or to phase-referencing
problems. \pmrvii\ appeared compact at all scales. The fifth one, namely
\pmrix, displays a bent one-sided radio jet, implying $\beta>0.29\pm0.05$ and
$\theta<73\pm3\degr$. Finally, \pmrx\ shows also a bent one-sided jet,
requiring $\beta>0.51\pm0.04$ and $\theta<60\pm3\degr$.

After a detailed analysis of our data, we show in Table~\ref{table:summary} a
summary of the results obtained after the VLA and optical observations
(Paper~I), and the EVN+MERLIN observations reported here. As can be seen, the
first two sources, \pmri\ and \pmriii, are promising microquasar candidates.
\pmriv\ is probably of thermal nature due to the highly inverted spectrum at
high radio frequencies, while \pmrvii\ is probably an extragalactic object due
to the extended nature of the optical counterpart. \pmrix\ shows properties
common to blazars, while \pmrx\ is an already identified quasar. We note that
\pmrix\ is bright enough at radio wavelengths to attempt an \ion{H}{i}
absorption experiment, that could allow to determine if this source is
galactic or not. In any case, optical spectroscopic observations of the first
five sources are in progress, to clearly unveil their galactic or
extragalactic nature.

\begin{acknowledgements}

We acknowledge R.~Porcas and W.~Alef for their useful comments and suggestions after reading through a draft version of this paper.
We acknowledge useful comments from L.~F. Rodr\'{\i}guez, the referee of this paper.
We are very grateful to S.~T. Garrington, M.~A. Garrett, D.~C. Gabuzda, and C.~Reynolds for their valuable help in the data reduction process.
This paper is based on observations with the 100-m telescope of the MPIfR (Max-Planck-Institut f\"ur Radioastronomie) at Effelsberg.
We thank the staff of the JIVE correlator and of the observing telescopes, especially A. Kraus for the single dish flux density measurements at the 100~m antenna in Effelsberg.
The European VLBI Network is a joint facility of European, Chinese and other radio astronomy institutes funded by their national research councils.
MERLIN is operated as a National Facility by the University of Manchester at Jodrell Bank Observatory on behalf of the UK Particle Physics \& Astronomy Research Council.
The EVN observations were carried out thanks to the TMR Access to Large-scale Facilities programme under contract No. ERBFMGECT950012.
Part of the data reduction was done at JIVE with the support of the European Community - Access to Research Infrastructure action of the Improving Human Potential Programme under contract No. HPRI-CT-1999-00045.
M.~R., J.~M.~P. and J.~M. acknowledge partial support by DGI of the Ministerio de Ciencia y Tecnolog\'{\i}a (Spain) under grant AYA2001-3092, as well as partial support by the European Regional Development Fund (ERDF/FEDER).
During this work, M.~R. has been supported by two fellowships from CIRIT (Generalitat de Catalunya, ref. 1998~BEAI~200293 and 1999~FI~00199).
J.~M. has been aided in this work by an Henri Chr\'etien International Research Grant administered by the American Astronomical Society, and has been partially supported by the Junta de Andaluc\'{\i}a.
This research has made use of the NASA's Astrophysics Data System Abstract Service, of the SIMBAD database, operated at CDS, Strasbourg, France, and of the NASA/IPAC Extragalactic Database (NED) which is operated by the Jet Propulsion Laboratory, California Institute of Technology, under contract with the National Aeronautics and Space Administration. 

\end{acknowledgements}

\end{document}